\documentclass[floatfix,aps,amsmath,nofootinbib,onecolumn,12pt]{revtex4}
\pdfoutput=1
\usepackage{graphicx}
\usepackage{epstopdf}
\usepackage{amsmath}
\usepackage{amsfonts}
\usepackage{amssymb}
\usepackage{color}
\usepackage{mathrsfs}
\usepackage{multirow}
\usepackage{bm}

\def\half{{1\over 2}}

\def\half{{1\over 2}}
\def\({\left(}
\def\){\right)}
\def\[{\left[}
\def\]{\right]}

\def\e{\begin{equation}}
\def\q{\end{equation}}
\def\m{\begin{eqnarray}}
\def\n{\end{eqnarray}}

\begin{document}

\title{Inflation model constraints from data released in 2015}

\author{Qing-Guo Huang \footnote{huangqg@itp.ac.cn}, Ke Wang \footnote{wangke@itp.ac.cn} and Sai Wang \footnote{wangsai@itp.ac.cn}}
\affiliation{Key Laboratory of Theoretical Physics, Institute of Theoretical Physics, Chinese Academy of Sciences, Beijing 100190, People's Republic of China}

\date{\today}

\begin{abstract}

We provide the latest constraints on the power spectra of both scalar and tensor perturbations from the CMB data (including \textit{Planck}~2015, BICEP2 \& \textit{Keck Array} experiments) and the new BAO scales from SDSS-III BOSS observation. We find that the inflation model with a convex potential is not favored and both the inflation model with a monomial potential and the natural inflation model are marginally disfavored at around $95\%$ confidence level. But both the Brane inflation model and the Starobinsky inflation model fit the data quite well.


\end{abstract}


\maketitle

\newpage

\section{Introduction}

In the past three decades, the inflation model \cite{Starobinsky:1980te,Guth:1980zm,Linde:1981mu,Albrecht:1982wi} was taken as the standard paradigm of the very early Universe. It not only resolves the flatness, horizon and monopole problems in the hot big bang model, but also generates the primordial density perturbations seeding the anisotropies in the cosmic microwave background (CMB) and the large-scale structure in our Universe. The simple inflation model predicts that the power spectra of both primordial scalar and tensor perturbations are adiabatic, Gaussian and nearly scale-invariant. In particular, the amplitude of primordial tensor perturbations is determined by the energy scale of inflation. Actually the primordial perturbations contribute to the anisotropies and polarizations of CMB as well as CMB lensing, which can be precisely measured by the ground-based or satellite CMB experiments, for example the \textit{Planck} satellite \cite{Adam:2015rua,Ade:2015lrj}. Therefore one can estimate the cosmological parameters and explore the nature of inflation by using the CMB data.

The excess of B-mode power over the base lensed-$\Lambda$CDM expectation detected by BICEP2 \cite{Ade:2014xna} can be explained by the polarized thermal dust, not the primordial gravitational waves \cite{Mortonson:2014bja,Flauger:2014qra}. Subtracting the contributions to the CMB B-mode from Galactic polarized dust measured by Planck collaboration in \cite{Adam:2014bub}, the tensor-to-scalar ratio $r\lesssim 0.1$ at $95\%$ confidence level (C.L.) was obtained in \cite{Cheng:2014pxa} which was confirmed by a joint analysis of B-mode polarization data of BICEP2/Keck Array and \textit{Planck} (BKP) in \cite{Ade:2015tva}. Furthermore, because of the tight constraint on the tensor-to-scalar ratio, the chaotic inflation model with a potential $V(\phi)\propto \phi^2$ was disfavored at more than $2\sigma$ C.L. in \cite{Cheng:2014pxa}. All of these results are confirmed by the \textit{Planck}~2015 results \cite{Ade:2015lrj} in which the scalar spectral index $n_s=0.968\pm0.006$ at $68\%$ C.L. and the tensor-to-scalar ratio $r_{0.002}<0.11$ at $95\%$ C.L. by fitting the \textit{Planck} TT,TE,EE+lowP+lensing combination (P15).

Recently BICEP2 \& \textit{Keck Array} CMB polarization experiments released the B-mode polarization data up to and including the 2014 observing season in \cite{Array:2015xqh}. This dataset includes the first \textit{Keck Array} B-mode data at $95$ GHz. The BICEP2 \& \textit{Keck Array} B-mode data (BK14) implies $r_{0.05}<0.09$ ($95\%$ C.L.). Combining with \textit{Planck}~2015 TT+lowP+lensing and some other external data, the upper bound on $r$ becomes $r_{0.05}<0.07$ ($95\%$ C.L.) \cite{Array:2015xqh} in the base $\Lambda$CDM+$r$ model. This constraint on $r$ is the strongest one to date, even though it is model-dependent in some sense.

In addition, the Baryon Acoustic Oscillation (BAO) data can significantly break the degeneracies between cosmological parameters. Recently, the BAO distance scale measurements were updated via an anisotropic analysis of BAO scale in the correlation function \cite{Cuesta:2015mqa} and power spectrum \cite{Gil-Marin:2015nqa} of the CMASS and LOWZ galaxy samples from Data Release 12 of the SDSS-III Baryon Oscillation Spectroscopic Survey (BOSS DR12). The total volume probed in DR12 has a $10\%$ increment from DR11 and the experimental uncertainty has been reduced. In this paper, the BAO data we will adopt include 6dFGS \cite{Beutler:2011hx}, MGS \cite{Ross:2014qpa}, BOSS DR12 CMASS \cite{Gil-Marin:2015nqa} and LOWZ \cite{Gil-Marin:2015nqa} (BAO15).

In this paper we will make a joint analysis of the recently released CMB and BAO data to constrain the cosmological parameters and the inflation models. Our paper is arranged as follows. In Sec.~\ref{dataanalysis}, we constrain the power spectra of both scalar and tensor perturbations by using the data combination of P15+BK14+BAO15. In Sec.~\ref{constraintsoninflation}, we will test several inflation models in two different methods. The summary and discussion are included in Sec.~\ref{summary}.

\section{Constraints on the power spectra of scalar and tensor perturbations}\label{dataanalysis}

In this section, we will make a joint analysis of P15+BK14+BAO15 dataset to constrain the cosmological parameters in the base $\Lambda$CDM+$r$ model and the other two extended models.

In general, the power spectra of primordial scalar and tensor perturbations can be parameterized as follows
\m
P_s(k)&=&A_s\left(\frac{k}{k_p}\right)^{n_s-1+\frac{1}{2}\frac{d n_s}{d\ln k}\left(\ln \frac{k}{k_p}\right)+\frac{1}{6}\frac{d^2 n_s}{d\ln k^2}\left(\ln \frac{k}{k_p}\right)^2}\ ,\\
P_t(k)&=&A_t\left(\frac{k}{k_p}\right)^{n_t}\ ,
\n
where $A_s$ and $A_t$ denote the amplitudes of scalar and tensor power spectra at the pivot scale $k_p$, $n_s$ and $n_t$ denote the spectral indices of scalar and tensor power spectra, $n_{\textrm{run}}\equiv d n_s/d \ln k$ and $n_{\textrm{run,run}}\equiv d^2 n_s/d \ln k^2$ denote the running and the running of running of scalar spectral index. In this paper, we set the pivot scale as $k_p=0.01~\textrm{Mpc}^{-1}$. The spectral index of tensor power spectrum is set as $n_t=-A_t/(8A_s)$ which is the consistency relation for the canonical single-field slow-roll inflation model \cite{Liddle:1992wi}. In principle, $n_t$ can be taken as a free parameter. However, $n_t$ cannot be well constrained by the current observations \cite{Cheng:2014ota,Huang:2015gka,Cabass:2015jwe}, even though $n_t\simeq 0$ is consistent with the current datasets. The measurement on $n_t$ is expected to be improved by some forthcoming experiments \cite{Huang:2015gca}.

In this paper we consider three cosmological models. The first one is the base $\Lambda$CDM+$r$ model in which there are seven free parameters: the baryon density today $(\Omega_b h^2)$, the cold dark matter density today $(\Omega_c h^2)$, the $100\times$ angular scale of the sound horizon at last-scattering ($100\theta_{\rm MC}$), the Thomson scattering optical depth due to the reionization $(\tau)$, the amplitude of scalar power spectrum $(A_s)$, the spectral index of scalar power spectrum $(n_s)$ and the tensor-to-scalar ratio $(r)$. The second one is the base $\Lambda$CDM+$r$+nrun model where the running of scalar spectral index ($d n_s/d \ln k$) is included as an additional free parameter. The third one is the base $\Lambda$CDM+$r$+nrun+nrunrun model in which an additional parameter, namely the running of running ($d^2 n_s/d \ln k^2$), is added.

The BAO data and $1\sigma$ uncertainty of BOSS DR12 LOWZ and CMASS samples are listed in Tab.~\ref{BAO}.
\begin{table*}[!htp]
\centering
\renewcommand{\arraystretch}{1.5}
\begin{tabular}{ccccc}
\hline\hline
$z_{eff}$     &BOSS DR12        &$H(z_{eff})r_{d}[10^3\textrm{km}\cdot s^{-1}]$     &$D_A(z_{eff})/r_{d}$    &$\rho_{D_A,H}$  \\
\hline
$0.32$                 &LOWZ        &$11.64\pm0.70$                            &$6.76\pm0.15$       &$0.35$    \\
\hline
$0.57$                 &CMASS       &$14.66\pm0.42$                            &$9.47\pm0.13$       &$0.54$    \\
\hline
\end{tabular}
\caption{The distance measurement from the anisotropic analysis of BAO scale in the CMASS and LOWZ galaxy samples released by SDSS-III BOSS DR12 \cite{Gil-Marin:2015nqa}.}
\label{BAO}
\end{table*}
Their consensus values are listed here, which are used in this paper. Here $z_{eff}$ denotes the effective redshift for CMASS and LOWZ samples, respectively. $H(z_{eff})$ and $D_A(z_{eff})$ are Hubble parameter and angular diameter distance at redshift $z_{eff}$, respectively. $r_d$ denotes the comoving sound horizon at the redshift of baryon drag epoch. In addition, $\rho_{D_A,H}$ is the normalized correlation between $D_A(z_{eff})$ and $H(z_{eff})$.

We can add a likelihood of the above BAO scales into our data analysis. The likelihood is given by
\begin{equation}
-2\ln \mathcal{L}=\sum_{k=1}^2\left(D^{data}-D^{model}\right)^{T}_k C_k^{-1}\left(D^{data}-D^{model}\right)_k\ ,
\end{equation}
where $k=1$ for LOWZ and $k=2$ for CMASS.
We denote $D^{data}_k\equiv\left(Hr_d,D_Ar_d\right)^{T}_k$ for the observational BAO data in Tab.~\ref{BAO}. Calculated by CAMB \cite{Lewis:1999bs,camb2012}, $D^{model}_k$ denote the corresponding theoretical predictions of cosmological models. We denote $C_k$ as the covariance matrix, namely
\begin{equation}
C_k=\left(\begin{array}{cc}
\sigma_{D_A}^2 & \rho_{D_A,H}\sigma_{D_A}\sigma_{H} \\
\rho_{D_A,H}\sigma_{D_A}\sigma_{H} & \sigma_{H}^2
\end{array}\right)_k
\end{equation}
where $\sigma_{D_A}$ and $\sigma_{H}$ respectively denote $1\sigma$ uncertainty of $Hr_d$ and $D_Ar_d$ in Tab.~\ref{BAO}.

In this paper, we use the Markov Chain Monte Carlo sampler (CosmoMC) \cite{Lewis:2002ah} to explore the space of cosmological parameters in each cosmological model.
Our results are summarized in Tab.~\ref{tab:constraints}, in which we list the constraints on all of the cosmological parameters as well as the best-fit $\chi^2$ for these three cosmological models.
\begin{table}[htp]
\scriptsize
\centering
\renewcommand{\arraystretch}{1.5}
\begin{tabular}{c|c|c|c}
 \hline
   $\textrm{Parameters}$ & $~~~~~~~~~\Lambda$CDM+$r~~~~~~~~~$ & $~~~~~\Lambda$CDM+$r$+nrun~~~~~ & $~\Lambda$CDM+$r$+nrun+nrunrun~ \\
\hline
$\Omega_bh^2$& $0.02227\pm0.00014$ &  $0.02230\pm0.00015$ &  $0.02223\pm0.00015$ \\
$\Omega_ch^2$& $0.1188\pm0.0010$ &  $0.1188\pm0.0010$ &  $0.1190\pm0.0010$ \\
$100\theta_{MC}$& $1.04091\pm0.00030$ &  $1.04091\pm0.00030$ &  $1.04092\pm0.00029$ \\
$\tau$& $0.067\pm0.012$ &  $0.067\pm0.012$ &  $0.072\pm0.013$ \\
$\ln(10^{10}A_s)$& $3.119\pm0.021$ &  $3.117\pm0.021$ &  $3.132\pm0.023$ \\
$n_s$& $0.9669\pm0.0040$ &  $0.9721\pm0.0108$ &  $0.9756\pm0.0111$ \\

$r_{0.01}~(95\%~\textrm{C.L.})$& $<0.0685$ &  $<0.0751$ &  $<0.0814$ \\
$d n_s/d\ln k$& - &  $-0.0035\pm0.0068$ &  $-0.0247\pm0.0148$ \\
$d^2 n_s/d\ln k^2$& - &  - &  $0.0211\pm0.0130$ \\
\hline
$\chi^2$& $13608.6$ &  $13608.6$ &  $13605.3$ \\
\hline
\end{tabular}
\caption{The $68\%$ limits on the cosmological parameters in three cosmological models from P15+BK14+BAO15 data combination. }
\label{tab:constraints}
\end{table}

\begin{figure}[htbp]
\begin{center}
\includegraphics[width=8 cm]{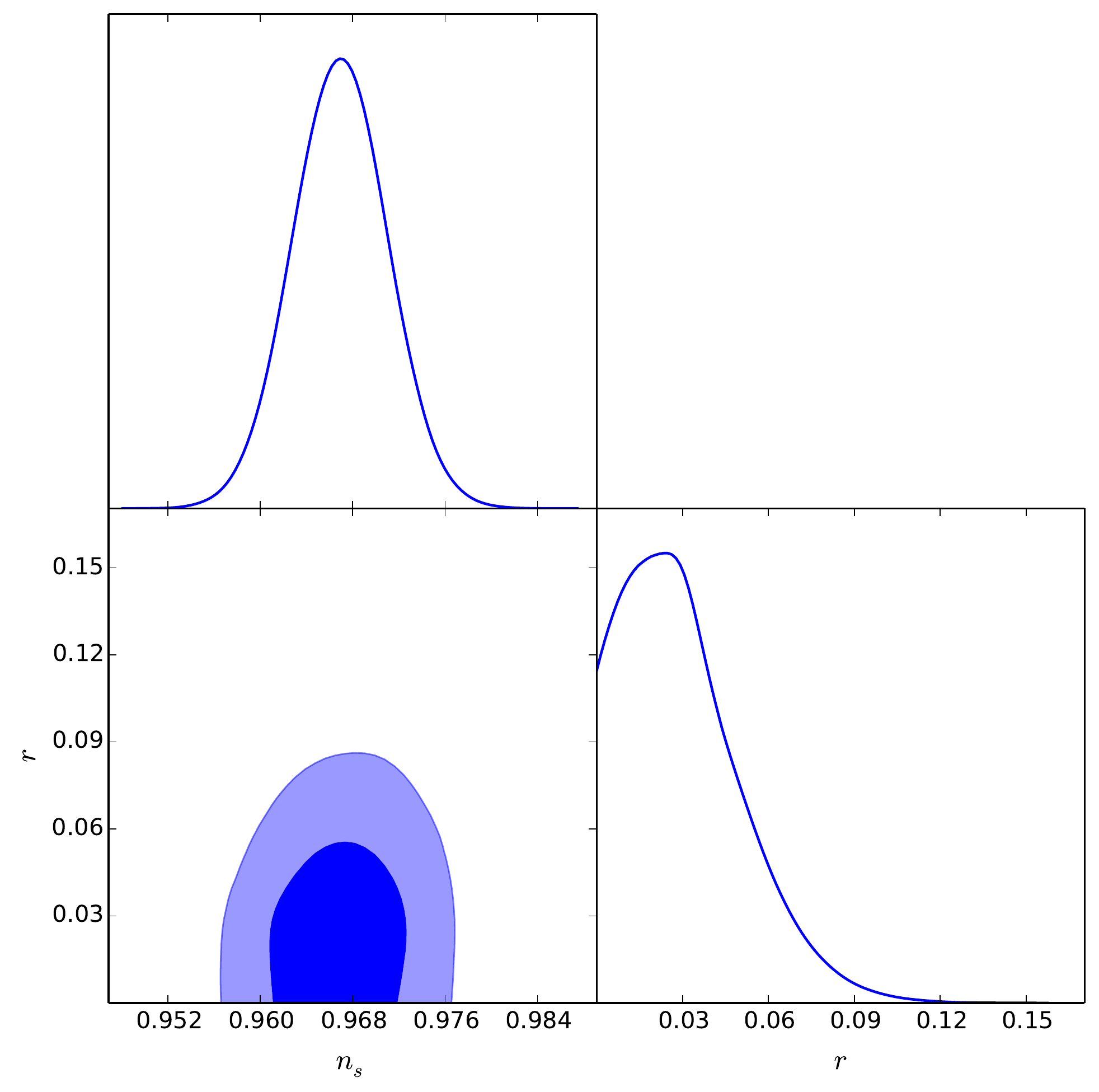}
\caption{The marginalized contour plots and the likelihood distributions of $r$ and $n_s$ in the base $\Lambda$CDM+$r$ model. }\label{fig:nsr1}
\end{center}
\end{figure}

The marginalized contour plots and the likelihood distributions of $r$ and $n_s$ in the base $\Lambda$CDM+$r$ model are depicted in Fig.~\ref{fig:nsr1}.
The constraints on $r$ and $n_s$ are given by
\m
&&r_{0.01}<0.0685~~(95\% ~\textrm{C.L.})\ ,\\
&&n_s=0.9669\pm 0.0040~~(68\% ~\textrm{C.L.})\ .
\n
The primordial scalar power spectrum deviates from the Harrison-Zel'dovich spectrum at more than $8\sigma$ C.L..

\begin{figure}[htbp]
\begin{center}
\includegraphics[width=12 cm]{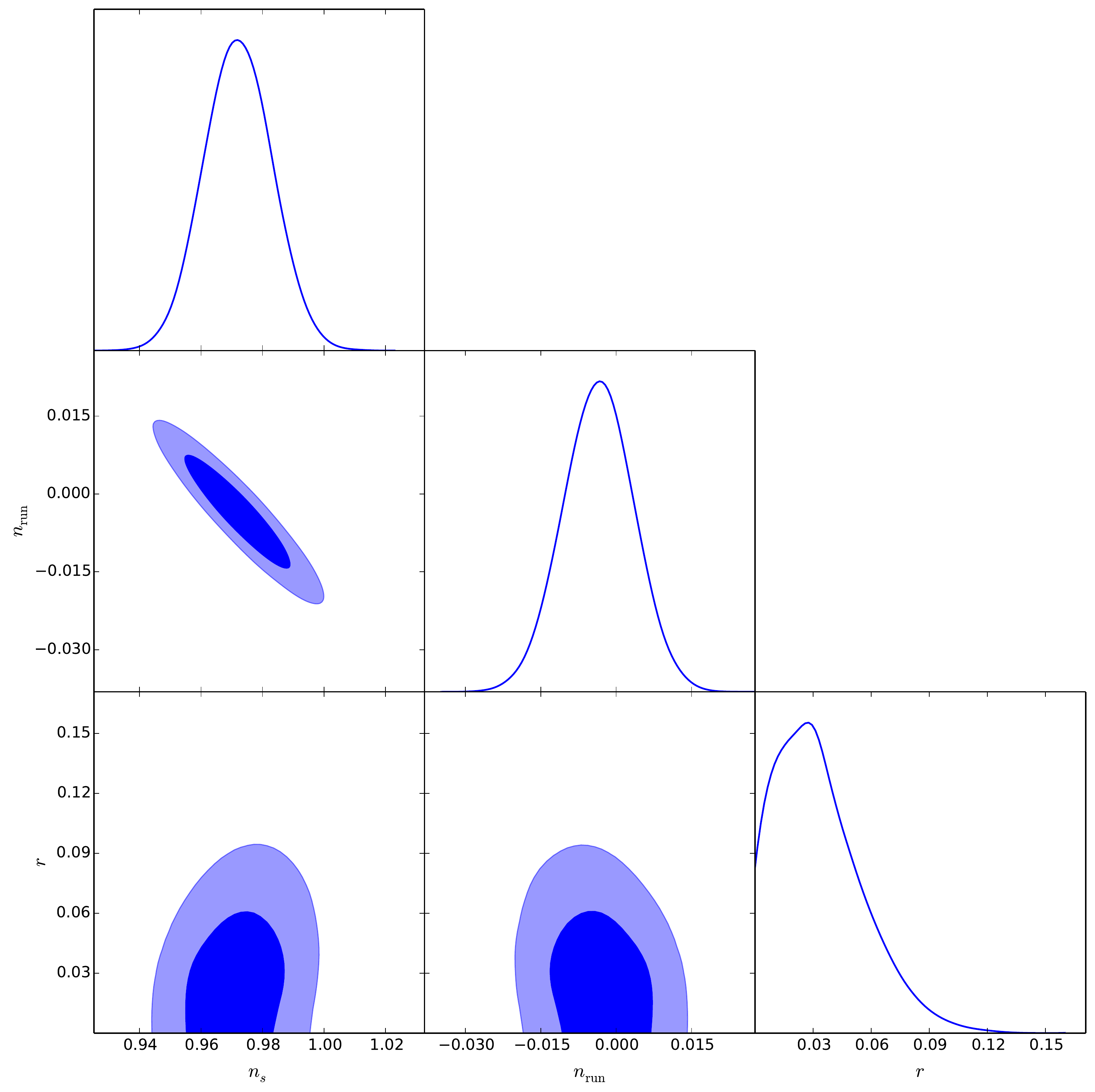}
\caption{The marginalized contour plots and the likelihood distributions of $r$, $n_s$ and $n_{\textrm{run}}\equiv d n_s/d\ln k$ in the base $\Lambda$CDM+$r$+nrun model. }\label{fig:nsr2}
\end{center}
\end{figure}

The marginalized contour plots and the likelihood distributions of $r$, $n_s$ and $d n_s/d\ln k$ in the base $\Lambda$CDM+$r$+nrun model show up in Fig.~\ref{fig:nsr2}.
We see that the constraints on $r$, $n_s$ and $d n_s/d\ln k$ read
\m
&&r_{0.01}<0.0751 ~~(95\% ~\textrm{C.L.})\ ,\\
&&n_s=0.9721\pm 0.0108~~(68\% ~\textrm{C.L.})\ ,\\
&&d n_s/d\ln k=-0.0035\pm0.0068 ~~(68\% ~\textrm{C.L.})\ .
\n
Even though there is one more parameter in the $\Lambda$CDM+$r$+nrun model, the best-fit $\chi^2$ is the same as that in the $\Lambda$CDM+$r$ model. We conclude that P15+BK14+BAO15 data do not prefer a non-zero running of scalar spectral index from the statistic point of view.

\begin{figure}[htbp]
\begin{center}
\includegraphics[width=16 cm]{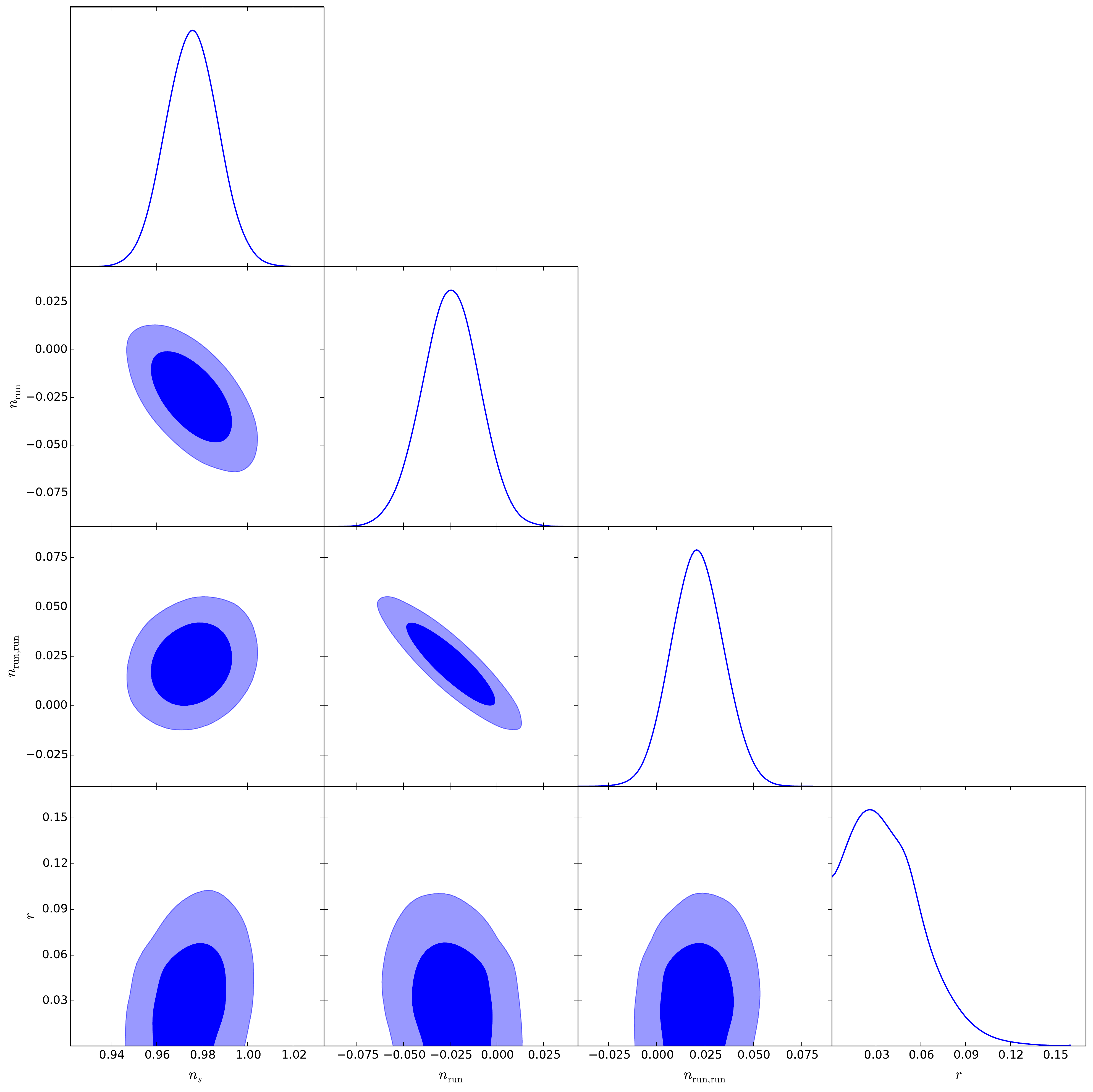}
\caption{The marginalized contour plots and the likelihood distributions of $r$, $n_s$, $n_{\textrm{run}}\equiv d n_s/d\ln k$ and $n_{\textrm{run,run}}\equiv d^2 n_s/d\ln k^2$ in the base $\Lambda$CDM+$r$+nrun+nrunrun model. }\label{fig:nsr3}
\end{center}
\end{figure}

Finally, the marginalized contour plots and the likelihood distributions of $r$, $n_s$, $d n_s/d\ln k$ and $d^2 n_s/d\ln k^2$ in the base $\Lambda$CDM+$r$+nrun+nrunrun model are illustrated in Fig.~\ref{fig:nsr3}.
We see that the constraints on $r$, $n_s$, $d n_s/d\ln k$ and $d^2 n_s/d\ln k^2$ are
\m
&&r_{0.01}<0.0814 ~~(95\% ~\textrm{C.L.})\ ,\\
&&n_s=0.9756\pm 0.0111~~(68\% ~\textrm{C.L.})\ ,\\
&&d n_s/d\ln k=-0.0247\pm0.0148~~(68\% ~\textrm{C.L.})\ ,\\
&&d^2 n_s/d\ln k^2=0.0211\pm0.0130 ~~(68\% ~\textrm{C.L.})\ .
\n
Our results show a slight preference for the negative running and a positive running of running of the scalar spectral index at the pivot scale $k_p=0.01$ Mpc$^{-1}$, and $\Delta \chi^2 \simeq -3.3$ compared to the base $\Lambda$CDM+$r$ model.

\section{Constraints on inflation}\label{constraintsoninflation}

The equations of motion for the canonical single-field slow-roll inflation take the form
\m
H^2&=&{1\over 3M_p^2} \[\half {\dot \phi}^2+V(\phi)\], \\
\ddot \phi &+&3H\dot \phi+V'(\phi)=0,
\n
where $M_p=1/\sqrt{8\pi G}$ is the reduced Planck energy scale, the dot and prime denote the derivative with respect to the cosmic time $t$ and the inflaton field $\phi$, respectively. The inflaton field slowly rolls down its potential if $\epsilon\ll 1$ and $|\eta|\ll 1$, where
\m
\epsilon&\equiv& {M_p^2\over 2} \({V'\over V}\)^2, \\
\eta&\equiv& M_p^2 {V''\over V}.
\n
In this limit, the equations of motion are simplified to
\m
H^2&\simeq&{V(\phi)\over 3M_p^2}, \\
3H\dot \phi&\simeq&-V'(\phi).
\n
From the dynamics of inflation, we have
\m
\eta=2\epsilon+\half {d\ln \epsilon\over dN}, \label{depsilon}
\n
where $N\equiv \int_t^{t_{\rm end}}H(t^\prime)dt^\prime$ is the number of e-folds before the end of inflation. The amplitudes of scalar and tensor power spectra are respectively given by
\m
P_s&=&{V\over 24\pi^2 M_p^2\epsilon}, \\
P_t&=&{2V\over 3\pi^2 M_p^2}.
\n
Therefore the tensor-to-scalar ratio and the spectral index of scalar power spectrum are related to the slow-roll parameters by
\m
r&=&16\epsilon, \\
n_s&=&1-6\epsilon+2\eta.
\n
See, for example, \cite{Huang:2014yaa} in detail.

\subsection{Selection of inflation models}

\begin{figure}[htbp]
\begin{center}
\includegraphics[width=12 cm]{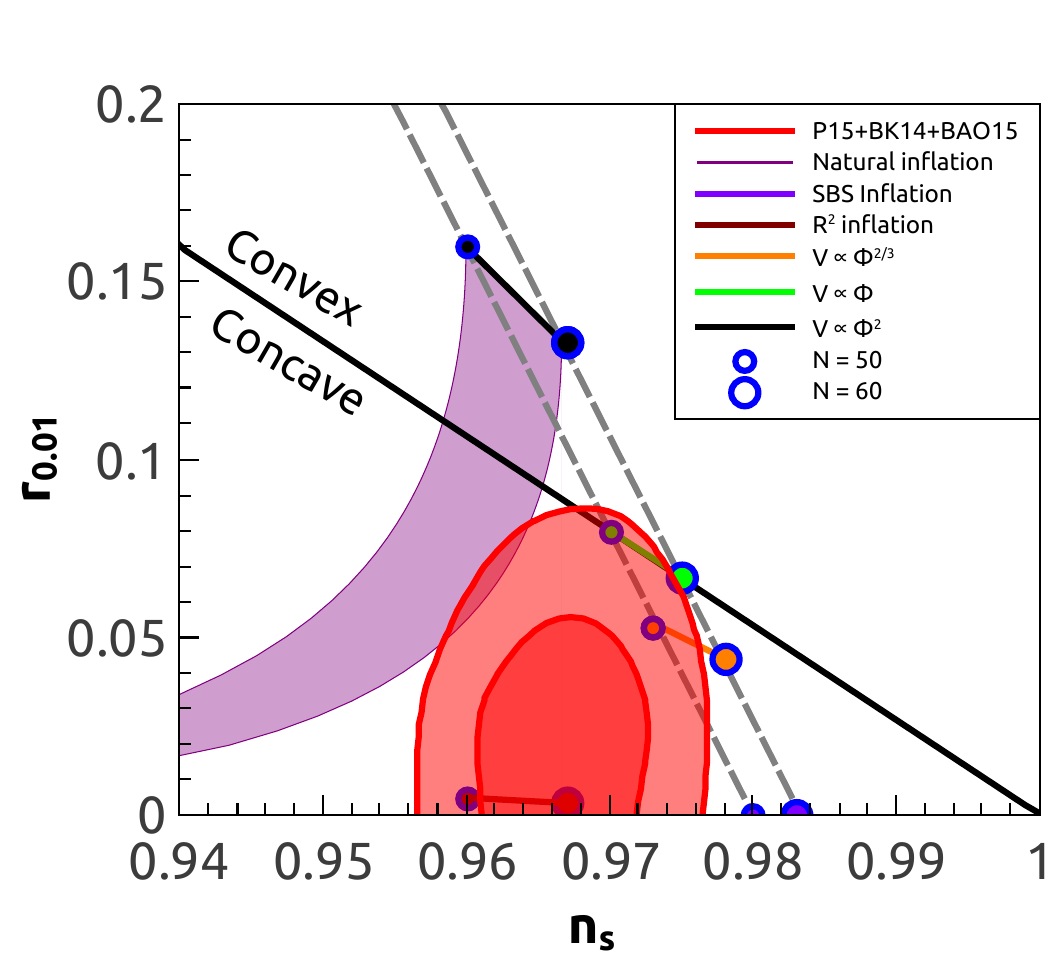}
\caption{Comparing the inflation models with the observational constraints. }\label{fig:nsr}
\end{center}
\end{figure}

There are a large number of inflation models in the market \cite{Martin:2013tda}. It is almost impossible to figure out a unique inflation model even when the cosmological parameters are measured very accurately, because the number of the cosmological parameters we can measure should be limited. In general, the simplicity is considered as a basic principle we should follow. In this subsection, we only take into account a few simple inflation models and compare them with the global fitting results given in the former section. Our main results are illustrated in Fig.~\ref{fig:nsr}. At first glance, we see that the inflation model with a concave potential is preferred at around $95\%$ C.L..

The inflation model with a monomial potential $V(\phi)\sim \phi^n$ \cite{Linde:1983gd} is the simplest class of inflation models, and is the prototype of chaotic inflation model. The predictions of this model are
\m
r&=&\frac{14n}{N}, \\
n_s&=&1-\frac{n+2}{2N}.
\n
Here $n$ is not necessarily an integer. Axion monodromy was supposed to achieve $V(\phi)\sim \phi^n$ inflation model in string theory. For example, $n=2/5$, $2/3$ in \cite{Silverstein:2008sg}, $n=1$ in \cite{McAllister:2008hb}, and the models with higher power in \cite{Marchesano:2014mla,McAllister:2014mpa}. For $50<N<60$, the predictions of $V(\phi)\sim \phi^n$ inflation model are illustrated in the region between two grey dashed lines in Fig.~\ref{fig:nsr}.
Unfortunately, we see that this class of inflation models are marginally disfavored at around $95\%$ C.L..

In the natural inflation model \cite{Freese:1990rb,Adams:1992bn}, the effective one-dimensional potential is given by  $V(\phi)=m^2f^2\left(1+\cos(\phi/f)\right)$ where $f$ denotes the decay constant. The natural inflation predicts
\m
r&=&\frac{8}{(f/M_p)^2}\frac{1+\cos\theta_N}{1-\cos\theta_N}, \\
n_s&=&1-\frac{1}{(f/M_p)^2}\frac{3+\cos\theta_N}{1-\cos\theta_N},
\n
where
\m
\cos\frac{\theta_N}{2}=\exp\left(-\frac{N}{2(f/M_p)^2}\right)\ .
\n
For $50<N<60$, the different decay constant corresponds to different predictions. See the purple shaded region in Fig.~\ref{fig:nsr}. Compared to the constraints from data, the natural inflation model is also marginally disfavored at $95\%$ C.L..

In the spontaneously broken SUSY (SBS) inflation model \cite{SUSY}, the potential of inflaton field takes the form of  $V(\phi)=V_0\left(1+c\ln\frac{\phi}{Q}\right)$ where $V_0$ is dominant and $c<<1$. This inflation model predicts
\m
r&\simeq&0, \\
n_s&=&1-\frac{1}{N}.
\n
It is also disfavored at more than $95\%$ C.L., because it predicts a large scalar spectral index. If the soft SUSY breaking term is taken into account, the scalar spectral index can shift to fit the data \cite{Rehman:2009nq,Pallis:2013dxa}.

In the Starobinsky inflation model \cite{Starobinsky:1980te}, the inflationary expansion of the Universe is driven by a higher derivative term in the action, namely $S=\frac{M_p^2}{2}\int d^4x\sqrt{-g}\left(R+\frac{R^2}{6M^2}\right)$ where $M$ denotes an energy scale. The tensor-to-scalar ratio and the scalar spectral index in Starobinsky inflation model are given by
\m
r&\simeq&\frac{12}{N^2}, \\
n_s&=&1-\frac{2}{N},
\n
in \cite{Mukhanov:1981xt,Starobinsky:1983zz}. Even though this inflation model can fit the data quite well, why the terms with higher powers of $R$ are all suppressed \cite{Huang:2013hsb} is still an open question.

\subsection{Constraints on typical inflation models}
\label{infcp}

In this subsection, we propose a new method to explore the space of inflation models. Similar to \cite{Huang:2007qz}, we parametrize the slow-roll parameter $\epsilon(N)$ in terms of e-folding number $N$ for the typical inflation model as follows
\m
\epsilon(N)={c/2\over (N+\Delta N)^p},
\label{epn}
\n
where $c(>0)$ and $p$ are two constants, and
\m
\Delta N=\({c\over 2}\)^{1/p}.
\n
Here we introduce $\Delta N$ to keep $N=0$ at the end of inflation, namely $\epsilon(N=0)=1$. See some extended investigations in \cite{Mukhanov:2013tua,Roest:2013fha,Garcia-Bellido:2014wfa,Gobbetti:2015cya}. Now the tensor-to-scalar ratio reads
\m
r={8c\over (N+\Delta N)^p}.
\n
From Eq.~(\ref{depsilon}), we obtain
\m
\eta&=&{c\over (N+\Delta N)^p}-{p\over 2(N+\Delta N)},
\n
and then
\m
n_s=1-{c\over (N+\Delta N)^p}-{p\over N+\Delta N}.
\n
Even though this parametrization can not cover all of the canonical single-field slow-roll inflation models, it can really cover many well-known inflation models. \\
$\bullet$ For the inflation model with $V(\phi)\sim \phi^n$, we have $p=1$ and $c=n/2$.
\\
$\bullet$ For the Starobinsky model, we have $p=2$ and $c=3/2$.
\\
$\bullet$ For the brane inflation model \cite{Dvali:1998pa} with $V(\phi)=V_0(1-(\mu/\phi)^{d-2})$, we have $p=2(d-1)/d$, where $d$ is the number of transverse dimensions. For example, the D3-Brane inflation in the KKLMMT setup predicts $p=5/3$ and $c\simeq 0$ \cite{Kachru:2003sx}.

\begin{figure}[htbp]
\begin{center}
\includegraphics[width=12 cm]{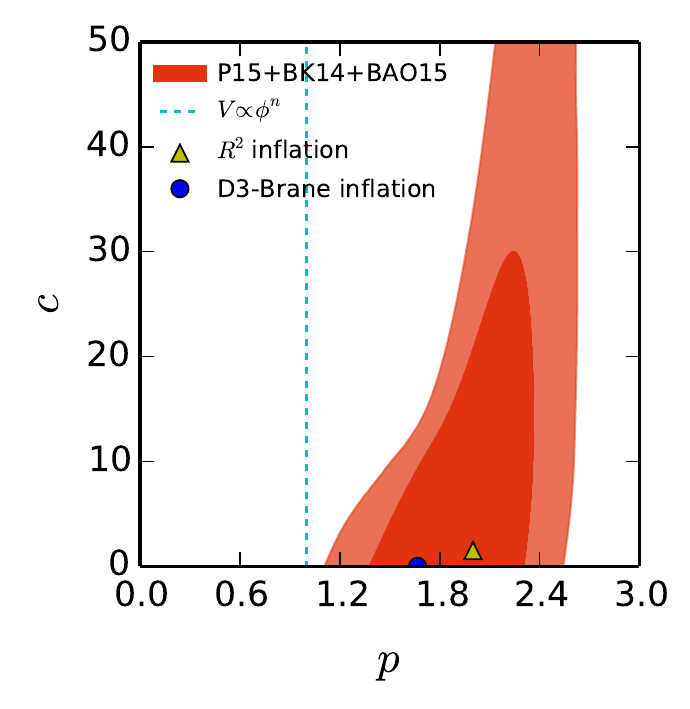}
\caption{Constraints on the space of inflation models from P15+BK14+BAO15. }\label{fig:cp}
\end{center}
\end{figure}

Up to now the exact e-folding number corresponds to the pivot scale $k_p$ is still unknown. Usually it is considered to be a number between 50 and 60. In this subsection, we take $N$ as a free parameter. Adopting the flat priors for $c\in [0,100]$, $p\in [0.5,3]$ and $N\in [50,60]$, globally fitting P15+BK14+BAO data combination and finally marginalizing over other free parameters in the base $\Lambda$CDM model and $N$, we obtain the constraints on the parameters $c$ and $p$, namely
\m
p&=&2.07_{-0.28}^{+0.32} \quad (68\%\ \textrm{C.L.}), \\
c&<&65.7 \quad (95\%\ \textrm{C.L.}).
\n
See Fig.~\ref{fig:cp} as well.
We see that the inflation model with a monomial potential corresponds to $p=1$ which is disfavored at more than $95\%$ C.L.. But both the Brane inflation model and the Starobinsky inflation model still fit the data quite well.

\section{Summary and Discussion}\label{summary}

In this paper, we constrained the cosmological parameters in the base $\Lambda$CDM+$r$ model and two extended models by jointly analyzing the data combination including \textit{Planck}~2015 data of CMB anisotropies and polarizations as well as CMB lensing, BICEP2 \& \textit{Keck Array} data of B-mode polarization up to and including 2014 observing season, and the anisotropic BAO distance scales from SDSS-III BOSS DR12 together with the isotropic 6dFGS and MGS BAO data.
Comparing with the predictions of inflation models, we find that the inflation model with a concave potential is preferred and both the inflation model with a monomial potential and the natural inflation model are marginally disfavored at $95\%$ C.L.. But both the Brane inflation model and the Starobinsky inflation model still give a good fit to the current data.


Even though there is no evidence for supporting a non-zero running of scalar spectral index, a positive running of running  in order of ${\cal O}(10^{-2})$ provides a slightly better fit to the data. Such a positive running of running can significantly relax the Lyth bound on the tensor-to-scalar ratio to $0.1$ \cite{Huang:2015xda}.

Before closing this paper, we also want to sketch the potential of inflaton field according to the current cosmological data. To summarize, the preferred inflation potential goes like that as follows
\m
V(\phi)=\Lambda_*-\delta V(\phi),
\n
where $\Lambda_*$ is an effective cosmological constant which determines the energy scale of inflation, and the dynamics of inflation is described by the subdominant term $\delta V(\phi)$. Since the concave shape of potential is preferred, $\delta V''(\phi)>0$. For example, for
\m
V(\phi)=\Lambda_* \[1-\({\phi\over \mu}\)^n\],
\n
where $\mu$ is an energy scale, $n$ is related to $p$ in Eq.~(\ref{epn}) by
\m
n={2(p-1)\over p-2}\quad \hbox{for}\ p>1\ \hbox{and}\ p\neq 2.
\n
Note that $p=2$ corresponds $V(\phi)=\Lambda_* \[1-\exp(-{\phi/ \mu})\]$. From Sec.~\ref{infcp}, we find $p>1$ which implies $n<0$ or $n>2$ and the potential is certainly concave. Now we still cannot distinguish the sign of $n$. We hope that the accurate experiments will tell us more detail about inflation in the near future.

\vspace{0.5 cm}
\noindent {\bf Acknowledgments}
We acknowledge the use of HPC Cluster of SKLTP/ITP-CAS. This work is supported by Top-Notch Young Talents Program of China and grants from NSFC (grant NO. 11322545, 11335012 and 11575271). QGH would also like to thank the participants of the advanced workshop ``Dark Energy and Fundamental Theory" supported by the Special Fund for Theoretical Physics from the National Natural Science Foundations of China (grant No. 11447613) for useful conversation.

\newpage


\end{document}